\documentclass[epj]{svjour}
\usepackage{graphics}
\begin{document}
\title{CPA density of states and conductivity in a 
double-exchange system containing impurities}
\titlerunning{CPA  in a double-exchange system}	
\author{Mark Auslender\inst{1}\thanks{\emph{e-mail:} 
marka@ee.bgu.ac.il}
\and Eugene Kogan\inst{2} \thanks{\emph{e-mail:} 
kogan@quantum.ph.biu.ac.il}}                     
%
%
\institute{Department of Electrical and Computer Engineering,
Ben-Gurion University of the Negev P.O.B. 653, Beer-Sheva 84105, Israel
\and Jack and Pearl Resnick Institute of Advanced Technology,
Department of Physics, Bar-Ilan University, Ramat-Gan 52900, Israel }
\authorrunning{Auslender and Kogan }
\date{Received: date / Revised version: date}
%
\abstract
{We study density of states and conductivity of the doped double-exchange
system, treating interaction of charge carriers both with the 
localized spins
and with the impurities in
the coherent potential approximation. It is shown that under appropriate
conditions
there is a gap  between  the
conduction band and the impurity band in  paramagnetic phase, while
the density of states is gapless in ferromagnetic phase. 
This can explain  metal-insulator transition frequently observed in manganites and
magnetic semiconductors.  Activated conductivity in the insulator phase is 
numerically calculated.
} 
\maketitle

\section{Introduction}

The recent rediscovery of colossal magnetoresistance (CMR) in doped Mn oxides
with perovskite structure R$_{1-x}$D$_x$MnO$_3$ (R is a rare-earth metal and D
is a divalent metal, typically Ba, Sr or Ca) \cite{helmolt}  has 
generated substantial interest in these materials \cite{ramirez}.  
The doping of parent material RMnO$_3$ by a divalent metal is the source of 
the holes responsible for the transport properties of these materials.
In addition, each divalent atom introduced, is the center of an impurity
potential. Many papers analyzed the influence of strong magnetic disorder, inherent in the
the CMR materials at finite temperature, upon the single-particle states and transport
properties. However, the interplay between the magnetic disorder and the doping-induced 
disorder was studied less. The  impurity potential plays double role. 
First, the potential fluctuations determine the transport at temperatures well 
below the ferromagnet (FM) - paramagnet (PM)  transition point $T_c$. Second, 
strong potential may pin the 
Fermi level  either in the conduction band tail 
(in the Anderson model 
of disorder \cite{sheng}), or in the emerging impurity band. 
The analysis of experimental data 
reveals strong relevance of the latter effect to metal-insulator transition (MIT) near $T_c$ 
both in magnetic semiconductors \cite{kog} and manganites \cite{bebenin}. 
However, to the best 
of our knowledge the impurity-band scenario in the double-exchange (DE) 
model was not  discussed yet. 
  
The present paper is devoted to the consideration of single-particle states and 
conductivity in impure DE system. Interaction of charge carriers both with the
localized spins and with the impurities is strong, so it is definitely not enough to 
limit ourselves with the  finite number of terms of
perturbation expansion. A simple but physically meaningful approximation,
allowing to sum up
infinite number of 
perturbation expansion terms  is  the
coherent potential approximation (CPA). Initially CPA was proposed
to treat potential disorder \cite{ziman}, but soon after it's appearance the
generalization to random spin system was developed \cite{kubo}. The CPA was also
used to describe diluted magnetic semiconductors \cite{takahashi}.
 
In the present paper we for the first time treat on equal footing the 
interactions of electrons with the core Mn spins and with the
doping impurities using 
the matrix generalization of the CPA. The concurrent action of 
potential disorder and temperature dependent spin disorder leads to a number of
interesting phenomena, in particular to the possibility of the opening of the
gap at the Fermi level with the increase of temperature and, hence, to
MIT transition.

\section{Hamiltonian and Theoretical Formulation}

We consider the DE model with the inclusion of the 
single-site impurity potential. In addition, as it is widely accepted, 
we apply the quasiclassical adiabatic approximation and consider each 
Mn spin as a static vector
with a fixed length $S$ (${\bf S}_i=S{\bf n}_i$, where ${\bf n}_i$
is a randomly oriented unit vector).
The Hamiltonian of the model in site representation is
\begin{eqnarray}
\label{ham}
\hat{H}_{ij}=t_{i-j}+ 
\delta_{ij}\left(\epsilon_i 
-J{\bf n}_i\cdot\hat{\bf \sigma}\right)=H_{kin}+V_{imp}+\hat{V}_{sd},
\end{eqnarray}
where $t_{i-j}$ is the electron hopping, $\epsilon_i$ is the 
random on-site energy, $J$ is the effective 
exchange 
coupling between a Mn core spin and a conduction electron and $
\hat{\bf \sigma}$ is the vector of the Pauli matrices. The hat above the
operator reminds that in one-particle representation it is a 
$2\times 2$ matrix in the spin space (we discard the
hat when the operator is a scalar matrix in the spin space).

We  present Hamiltonian as
\begin{equation}
\hat{H}=H_{kin}+\hat{\Sigma}+V_{imp}+\hat{V}_{sd}-\hat{\Sigma}=
\hat{H}_0+\hat{V}
\end{equation}
(the site independent self-energy $\hat{\Sigma}(E)$ is to be determined later),
and construct a perturbation theory with respect to random potential 
$\hat{V}=V_{imp}+\hat{V}_{sd}-\hat{\Sigma}$.
 To do this let us
introduce  the $T$-matrix as the solution of the equation
\begin{equation}
\hat{T}=\hat{V}+\hat{V}\hat{G}_0\hat{T},
\end{equation}
where
\begin{equation}
\label{efes}
\hat{G}_0=\frac{1}{E-\hat{H}_0}.
\end{equation}
For the exact Green function we get
\begin{equation}
\label{green}
\hat{G}=\hat{G}_0+\hat{G}_0\hat{T}\hat{G}_0.
\end{equation}
The coherent potential approximation (CPA)
is expressed by the equation
\begin{equation}
\label{g}
\left\langle\hat{G}\right\rangle=\hat{G}_0.
\end{equation}
This equation can also be presented as
\begin{equation}
\label{t}
\left\langle\hat{T}_i\right\rangle=0,
\end{equation}
where $\hat{T}_i$ is the solution of the equation
\begin{equation}
\hat{T}_i=\hat{V}_i+\hat{V}_i
\hat{g}(E-\hat{\Sigma})\hat{T}_i,
\end{equation}
and
\begin{equation}
\label{locator}
g(E)=\left(G_0(E)\right)_{ii}=
\int\frac{N_0(\varepsilon)}{E-\varepsilon}d\varepsilon,
\end{equation} 
where $N_0(\varepsilon)$ is the bare density of states.
The averaging in Eqs. (\ref{g},\ref{t}) should be performed both with respect to
random orientations of core spins and with respect to random on-site energies.
We obtained, in fact, the algebraic equation for the $2\times 2$ matrix $\hat{\Sigma}$ 
\begin{equation}
\label{gencpa}
\left\langle\left[1-\hat{V}_i\hat{g}(E-\hat{\Sigma})\right]^{-1}
\hat{V}_i\right\rangle=0.
\end{equation}
This equation takes into account scattering both due to randomness of the core
spins, and due to the impurities. 
 If the impurity potential is negligible
($V=0$) this equation   coincides with the Eq.(20) of Ref.
\cite{furukawa} obtained in the dynamical mean field approximation (and also
 with those obtained for the
Falikov-Kimball model \cite{moller,DMFA}).

In the reference frame where the $z$ axis is directed along the magnetization, 
$\hat{\Sigma}$ 
is diagonal, and Eq.(\ref{gencpa}) reduces to the system of two equations for 
its diagonal 
matrix elements $\Sigma_{\sigma}(E)$ ($\sigma = \uparrow,\downarrow$). 
The equations acquire especially simple form at two extreme
particular cases,which we will analyze:

\noindent (a) $T=0$. The magnetic state is coherent FM with $n_i^z=1$, and 
 Eq. (\ref{gencpa}) takes the form 
\begin{equation}
\left\langle \frac{\epsilon_i \mp J-\Sigma _{\uparrow ,\downarrow }}
{1-\left( \epsilon_i \mp J-\Sigma _{\uparrow ,\downarrow }\right)
g(E-\Sigma_{\uparrow ,\downarrow })}\right\rangle =0  \label{ztcpa}.
\end{equation}

\noindent (b) $T\geq T_{c}$ and zero magnetic field. 
The magnetic state is isotropic PM with 
$\langle {\bf n_{i}}\rangle =0$, which leads to 
$\Sigma _{\downarrow }=\Sigma_{\uparrow }=\Sigma $, and 
 Eq. (\ref{gencpa}) takes the form.
\begin{eqnarray}
\label{htcpa}
\left\langle\frac{\epsilon_i +J-\Sigma }{1-\left( \epsilon_i +J-\Sigma\right)
g(E-\Sigma)}\right\rangle \nonumber\\+
\left\langle\frac{\epsilon_i -J-\Sigma }
{1-\left(\epsilon_i -J-\Sigma \right) g(E-\Sigma)}\right \rangle = 0.  
\end{eqnarray}

We will solve the equations Eqs.(\ref{ztcpa}) and (\ref{htcpa}) in
the strong Hund coupling limit 
($J\rightarrow \infty $). 
In this limit we obtain two decoupled spin sub-bands. The equation for the upper
sub-band, after shifting the energy by $-J$,  for both cases (a) and (b) can be written
down in unified form 
\begin{equation}
\left\langle \frac{1}{1-\left( \epsilon_i -\Sigma \right)
g( E -\Sigma) }\right\rangle =\alpha,
\label{infJcpa}
\end{equation}
where $\alpha =1$ for $T=0$, $\alpha =2$ for $T\geq T_{c}$. 
In the model of substitutional
disorder ($\epsilon_i = 0$ with probability $x$, and $\epsilon_i =V$ with probability 
$1-x$),  
Eq.(\ref{infJcpa}) takes the form 
\begin{equation}
\frac{1-x}{1+\Sigma g\left(E - \Sigma \right) }+
\frac{x}{1+\left(\Sigma -V\right) g\left( E -\Sigma\right) }=\alpha.  
\label{infJcpasd}
\end{equation}

\section{The CPA equations for semi-circular bare density of states}

We consider semi-circular (SC) bare DOS 
\begin{equation}
N_{0}( \varepsilon) =\frac{4}{\pi W}
\sqrt{1-\left( \frac{2\varepsilon}{W}\right) ^{2}},
\label{scdos}
\end{equation}
at $\left| \varepsilon\right| \leq W/2$ and $N_{0}(\varepsilon) =0$ 
otherwise, for
which 
\begin{equation}
g(E) =\frac{4}{W}\left[\frac{2E}{W}-\sqrt{\left( \frac{2E}{W}\right)^{2}-1}\right].  
\label{scgrfun}
\end{equation}
Let us introduce the following normalized quantities 
\begin{eqnarray}
\lambda=\frac{\Sigma}{W},\;\omega =\frac{E}{W},\;v=\frac{V}{W}
\label{dimlessgf}
\end{eqnarray}
After simple algebra we  obtain from Eq. (\ref{infJcpasd}) 
the cubic equation with respect to 
\begin{eqnarray}
\gamma \equiv Wg\left(E -\Sigma \right) =8\left[ \omega -\lambda -
\sqrt{\left( \omega -\lambda \right) ^{2}-1/4}\right],  
\end{eqnarray}
in the form
\begin{eqnarray}
\gamma ^{3}+16\left( v-2\omega \right) \gamma ^{2}+16
\left[\frac{1}{\alpha} -16\omega \left( v-\omega \right) \right]
\gamma\nonumber\\ 
 -256\frac{\omega}{\alpha}
+256\left( 1-x\right) \frac{v}{\alpha} =0  \label{cubicgam}.
\end{eqnarray}
The  number of electrons per cite $n$ is given by
\begin{equation}
\label{fermi}
n=\int_{-\infty}^{\infty}f(E)N(E)dE,
\end{equation}
where $f(E)$ is the Fermi distribution function, and 
\begin{equation}
N(E)=\frac{\alpha}{W\pi}\mbox{Im}\;\gamma
\end{equation}
is the actual density of states.
To define the position of $\mu$, the Fermi level, 
we must impose the relation
between $n$ and $x$; the simplest assumption appropriate for manganites is the
equation $n=1-x$.

\section{Conductivity in CPA}

For a disordered one-electron system the static conductivity is given by

\begin{eqnarray}
\rho^{-1} =\frac{e^{2}\pi \hbar }{\rm V}
\int \left(-\frac{\partial f}{\partial E}\right)\nonumber\\
\cdot\left\langle \mbox{Tr}\left[ \hat{v}_{\alpha }\delta \left(E-\hat{H}\right) 
\hat{v}_{\alpha }\delta \left( E-\hat{H}\right) \right]
\right\rangle dE  \label{kubogrin},
\end{eqnarray}
where ${\rm V} $ is the volume and $\hat{v}_{a}$ is a Cartesian component of 
the velocity operator. To
obtain the conductivity in CPA let us express operator delta-function as
follows
\begin{equation}
\delta \left( E-\hat{H}\right) =\frac{1}{2\pi i}\left[ \hat{G}%
(E_{-})-\hat{G}(E_{+})\right].
\label{specop}
\end{equation}
Using Eq. (\ref{green}) and Eq. (\ref{efes}) in Bloch representation
\begin{equation}
\left\langle {\bf k}\sigma \left| \hat{G}_{0}(E)\right| {\bf k}%
^{\prime }\sigma ^{\prime }\right\rangle =\frac{\delta _{\mathbf{k,k}%
^{\prime }}\delta _{\sigma,\sigma^{\prime }}}{E
-\varepsilon_{\mathbf{k}}-\Sigma _{\sigma }\left( E\right) },
\label{grinfun}
\end{equation}
we get 
\begin{eqnarray}
\left\langle \mbox{Tr}\left[ \hat{v}_{a}\delta \left( E-\hat{H}%
\right) \hat{v}_{a}\delta \left( E-\hat{H}\right) \right]
\right\rangle \nonumber\\
=\sum_{\mathbf{k},\sigma }v_{\mathbf{k}\alpha }^{2}\left[
A_{\sigma }\left( \varepsilon_{\mathbf{k}},E\right) \right] ^{2}+
O(\left\langle \hat{T}\hat{T}\right\rangle), 
\end{eqnarray}
where 
\begin{equation}
A_{\sigma }\left( \varepsilon ,E\right) 
 =\frac{1}{\pi}\frac{\mbox{Im}\Sigma _{\sigma
}\left( E\right) }{\left[ E-\varepsilon -\mbox{Re}\Sigma _{\sigma }\left(
E\right) \right] ^{2}+\left[ \mbox{Im}\Sigma _{\sigma }\left( E\right)
\right] ^{2}}  \label{speccpa}
\end{equation}
is the one-particle spectral weight function. On account of the locality
of $T$-matrix the second term in the trace is equal to 
\begin{eqnarray}
\sum_{s,s^{^{\prime }}=\pm;\; 
\mathbf{k,k}^{\prime },\sigma ,\sigma^{\prime }}
ss^{\prime }v_{\mathbf{k}\alpha }v_{\mathbf{k}^{\prime }\alpha
}G_{\sigma }\left( \varepsilon_{\mathbf{k}},E_{s}\right) G_{\sigma ^{\prime
}}\left( \varepsilon_{\mathbf{k}^{\prime }},E_{s}\right)\nonumber\\
\times G_{\sigma ^{\prime }}\left( \varepsilon_{\mathbf{k}^{\prime
}},E_{s^{\prime }}\right) G_{\sigma }\left( \varepsilon_{\mathbf{k}%
},E_{s^{\prime }}\right) \nonumber\\
\times \left\langle T_{\sigma \sigma ^{\prime }}
({\bf k}-{\bf k}^{\prime },E_{s})T_{\sigma ^{\prime }\sigma }
({\bf k}^{\prime }-{\bf k},E_{s^{\prime }})\right\rangle. 
\end{eqnarray}
Since in CPA $\left\langle T_{\sigma \sigma ^{\prime }}
({\bf k}-{\bf k}^{\prime },E_{s})T_{\sigma ^{\prime }\sigma }
({\bf k}^{\prime }-{\bf k},E_{s^{\prime }})\right\rangle$ does not depend on 
$\mathbf{k}$ and $\mathbf{k}^{\prime}$
and $v_{-\mathbf{k}\alpha}=-v_{\mathbf{k}\alpha}$ the above expression is identically
zero \cite{velic,khurana}. 
 Thus, finally
\begin{eqnarray}
\rho^{-1} =\frac{e^{2}\pi \hbar }{ v }\int 
\int \left(-\frac{\partial f}{\partial E}\right)\nonumber\\
\cdot v_{\alpha }^{2}\left( \varepsilon \right) N_{0}\left( \varepsilon \right)
\sum_{\sigma }\left[ A_{\sigma }\left( \varepsilon ,E\right) \right]
^{2}dEd\varepsilon,   
\label{cpacond}
\end{eqnarray}
where $v$ is the unit cell volume and by definition
\begin{equation}
v_{\alpha}^{2}\left( \varepsilon \right) N_{0}\left( \varepsilon \right)
=\frac{1}{N\hbar ^{2}}
\sum_{\mathbf{k}}\left( \frac{\partial \varepsilon_{\mathbf{k}}}
{\partial k_{\alpha }}\right) ^{2}\delta \left( \varepsilon -\varepsilon
_{\mathbf{k}}\right). 
\label{msqvel}
\end{equation}

Let us assume nearest-neighbor tight binding spectrum on simple $d$%
-hypercubic lattice ($v =a^{d}$) 
\begin{eqnarray}
\varepsilon_{\mathbf{k}} = -t\sum_{\alpha =1}^{d}\cos ak_{\alpha },  \nonumber \\
v_{\alpha}^{2}\left(\varepsilon \right) N_{0}\left( \varepsilon \right)
 =-\frac{a^{2}}{d\hbar
^{2}}\int_{-\infty }^{\varepsilon}\epsilon N_{0}\left( \epsilon \right)
d\epsilon.   
\label{nnres}
\end{eqnarray}
In SC DOS model
\begin{eqnarray}
v_{\alpha }^{2}\left( \varepsilon \right) N_{0}\left( \varepsilon \right) =-
\frac{4}{\pi W}\frac{a^{2}}{d\hbar ^{2}}\int_{-W/2}^{\varepsilon }z\sqrt{
1-\left( \frac{2z}{W}\right) ^{2}}dz\nonumber\\
=\frac{1}{3}\frac{W}{\pi }\frac{a^{2}}{%
d\hbar ^{2}}\left( 1-\frac{4\varepsilon ^{2}}{W^{2}}\right) ^{3/2}.
\end{eqnarray}
Substituting this result into Eq.(\ref{cpacond}) we obtain
\begin{equation}
\label{resist}
\rho ^{-1}=\sigma _{0}\int \left(-\frac{\partial f}{\partial E}\right)
\Lambda \left(E\right) dE, 
\end{equation}
with
\begin{equation}
\sigma _{0}=\frac{e^{2}}{2\pi da^{d-2}\hbar }
\end{equation}
being the Mott minimal metallic conductivity, and 
\begin{eqnarray}
\Lambda \left( E\right)  =\frac{2W}{3\pi}\int_{-W/2}^{W/2} \left(
1-\frac{4\varepsilon ^{2}}{W^{2}}\right) ^{3/2}\nonumber\\
\sum_{\sigma}\left\{ \mbox{Im}\left[ \frac{1}
{E-\varepsilon -\Sigma _{\sigma }\left( E\right) }\right] \right\}
 ^{2}d\varepsilon. 
\end{eqnarray}
For the strong Hund coupling  we obtain
\begin{eqnarray}
\label{econd}
\Lambda \left( E\right)  =
=\frac{4}{3\pi}\int_{-1}^{1}\left( 1-x^{2}\right) ^{3/2}
\left\{ \mbox{Im}\left[ \frac{1}{x-z}\right] \right\} ^{2}dx,
\end{eqnarray}
where $z = 2\left( \omega - \lambda \right)$.
Eqs. (\ref{resist}), (\ref{econd})
give the conductivity in the framework of bare SC DOS model 
 for arbitrary hole concentration $x$ 
and impurity potential strength $V$.

\section{Influence of the impurity potential}

First, consider density of states.
It is known, that within CPA for every $x$ there exists critical value of potential-to-bandwidth 
ratio $v_c(x)$ such that at $v > v_c(x)$ the separate impurity band splits off the conduction band 
(that is a gap opens in $N(E)$).  In
 our approach
we get two different curves $v_c(x,T=0)$ and $v_c(x,T \geq T_c)$, which present boundaries of 
metal-insulator and metal-semiconductor 'phase diagrams', respectively, 
in $(v,x)$ plane. Due to 
effect of  magnetic disorder it appears that $v_c(x,T=0) > v_c(x,T \geq T_c)$. 

For a typical concentration $x=0.2$
$v_{c}(0.2, T=0) \approx 0.49$ and $v_{c}(0.2, T\geq T_c) \approx 0.35$. 
So if we choose 
$v = 0.4$ both $N(E)$ and $\Lambda (E)$ must be gapless at $T=0$ but do have a gap at $T\geq T_c$. 
Numerical calculations of the DOS  performed at $T=0$ and $T\geq T_c$ for the above 
$x$ and $v$ clearly demonstrate FM-PM transition induced band 
splitting (Fig.1).
 
\begin{figure}
\resizebox{0.4\textwidth}{!}{%
 \includegraphics{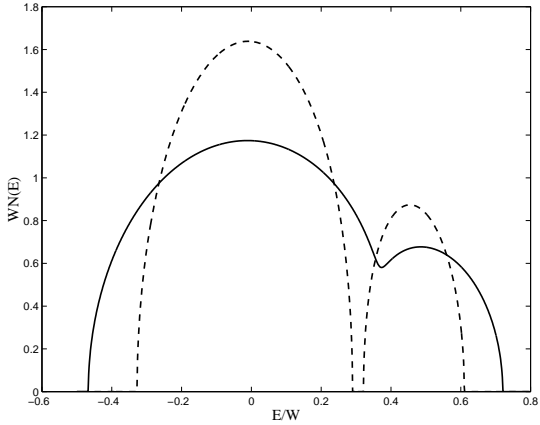}
}
\caption{DOS at $T=0$ (full line) and high-temperature (dashed line) 
in the units of $W^{-1}$ for $x=0.2$ and $V/W=0.4$}
\label{fig:1}       
\end{figure}

Now address the question of conductivity.
Consider first the position of $\mu$ and conductivity at $T=0$. 
We get from Eq.(\ref{fermi}) 
$\mu(T=0)=0.3662W$.  Note that $\mu(T=0)$ lies on the neck connecting 
conduction band and impurity states derived parts of the band. As a result,
the residual conductivity (Eq.(\ref{resist})) 
$\rho^{-1}(T=0)=0.6163\sigma_0$ is less then the Mott limit.

\begin{figure}
\resizebox{0.4\textwidth}{!}{%
 \includegraphics{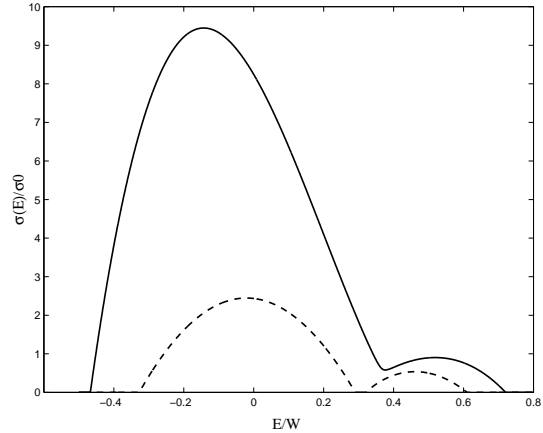}
}
\caption{Energy-dependent conductivity at $T=0$ (full line) 
and high-temperature
(dashed line) in the units of $\sigma_0$ for $x=0.2$ and $V/W=0.4$}
\label{fig:2}       
\end{figure}

At $T\geq T_c$ DOS and $\Lambda(E)$ 
have the same gap $\Delta=0.031W$ (see Figs.1,2), so $\mu(T\geq T_c)$ must lie 
in the gap. 
Thus,  the model describes a bad metal at
$T=0$, and a semiconductor  
at $T \geq T_c$. The transition between two types of
conduction (FM-PM 
transition induced MIT) should occur at some temperature below $T_c$. Such a
picture agrees with the 
recent photoemission experiments showing drastic decrease of DOS at the Fermi
level \cite{photo} 
as temperature increases towards $T_c$. 

It is checked numerically that DOS displays square-root like behavior near the top of the 
conduction band $E_{c}$ 
\begin{eqnarray}
N(E) \approx n_c\sqrt{E_{c}-E},
\end{eqnarray}
and the bottom of the impurity band $E_{i}$
\begin{eqnarray}
N(E) \approx n_i\sqrt{E-E_{i}}.
\end{eqnarray}
Unlike DOS $\Lambda (E)$ behaves {\it linearly} near the band edges
\begin{eqnarray}
\Lambda(E) \approx W^{-1}\lambda_c(E_{c}-E),\;\;\;\mbox{for}\; E<E_{c};\nonumber\\ 
\Lambda(E) \approx W^{-1}\lambda_i(E-E_{i}), \;\;\;\mbox{for}\; E>E_{i}.
\end{eqnarray}
The assumption $T<\Delta$
allows us to explicitly obtain $\mu(T\geq T_c)$.
Calculating integrals in 
Eq.(\ref{fermi}) with exponential accuracy we obtain
\begin{equation}
\mu \approx \frac{1}{2}\left( E_{c}+E_{i}+T\ln\frac{n_c}{n_i}\right).
\end{equation}
The integral in Eq.(\ref{resist}), calculated with the same accuracy, 
leads to activation law for conductivity with {\it linear} temperature 
pre-exponent
\begin{equation}
\rho^{-1} \approx \sigma_0 \frac{BT}{W}\exp\left(-\frac{E_A}{T}\right),
\end{equation}
where $E_A=\Delta/2\approx 0.015W$ and $B$ is the following numerical constant
\begin{equation}
B=\lambda_c \sqrt{\frac{n_i}{n_c}}+ \lambda_i \sqrt{\frac{n_c}{n_i}}\approx 22
\end{equation}
for the parameters considered.

Low values of conductivity obtained for the case of spin disorder 
are
an indication of the possibility of Anderson localization \cite{kog,li,deph}, 
which
CPA is incapable of accessing.  But the present results complement and support
the localization based approach. In fact, the results of Ref. \cite{kog} 
were obtained under the assumption of the Fermi level pinning, which is now
explained as being due to strong electron-impurities interaction (and the
impurity band formation). 

In another aspect, the model considered may also explain low-temperature MIT observed in initially metallic manganites 
R$_{1-x}$D$_x$MnO$_3$ upon substitution of R by isovalent atoms (e.g. La by Y \cite{barman}). 
One may speculate that the substitution forms a deep impurity band which can capture holes 
in R$_{1-x}$D$_x$MnO$_3$.

\section{Conclusion}
To conclude, we derived CPA equations for the one-electron Green function and 
conductivity of 
DE system containing impurities. The equations were solved for
the SC bare DOS and  substitutional disorder model. 
It was shown that
if the electron-impurity interaction is strong enough,   
 there is a gap  between  the
conduction band and the impurity band in  PM phase, 
the density of states being gapless in FM phase.  
Under appropriate doping conditions the chemical potential is pinned inside the
gap.  This can explain  metal-insulator transition observed in manganites and
magnetic semiconductors. 

\section{Acknowledgments}

This research was supported by the Israeli Science Foundation administered
by the Israel Academy of Sciences and Humanities.

\end{document}